\documentclass[aps,prl,twocolumn,superscriptaddress,showpacs,letterpaper,10pt]{revtex4-1}

\usepackage{graphicx}
\usepackage{xspace}
\usepackage{amsmath}
\usepackage{color}
\usepackage{hyperref}
\hypersetup{colorlinks=true,linkcolor=blue,citecolor=blue,urlcolor=blue}
\usepackage{txfonts}

\newcommand{\Zeff}{Z_\text{eff}}
\newcommand{\ZeffDir}{Z_\text{eff}^\text{(dir)}}
\newcommand{\ZeffRes}{Z_\text{eff}^\text{(res)}}
\newcommand{\ZeffMRA}{Z_\text{eff}^\text{(mra)}}
\newcommand{\ZeffTot}{Z_\text{eff}^\text{(tot)}}

\begin{document}

\title{Vibrational Feshbach Resonances Mediated by Nondipole Positron-Molecule Interactions}

\author{M.~R.~Natisin}
\thanks{Contact Information: michael.natisin.ctr@us.af.mil. Now at the Air Force Research Laboratory, Edwards AFB, CA 93524}
\affiliation{Department of Physics, University of California, San Diego, La Jolla CA 92093}
\author{J.~R.~Danielson}
\thanks{Contact information: jrdanielson@ucsd.edu}
\affiliation{Department of Physics, University of California, San Diego, La Jolla CA 92093}
\author{G. F. Gribakin}
\affiliation{Department of Applied Mathematics and Theoretical Physics, Queen's University, Belfast BT7 1NN, Northern Ireland, UK}
\author{A. R. Swann}
\affiliation{Department of Applied Mathematics and Theoretical Physics, Queen's University, Belfast BT7 1NN, Northern Ireland, UK}
\author{C.~M.~Surko}
\affiliation{Department of Physics, University of California, San Diego, La Jolla CA 92093}

\date{\today}

\begin{abstract}
Measurements of energy-resolved positron-molecule annihilation show the existence of positron binding and vibrational Feshbach resonances. The existing theory describes this phenomenon successfully for the case of infrared-active vibrational modes which allow dipole coupling between the incident positron and the vibrational motion. Presented here are measurements of positron-molecule annihilation made using a recently developed cryogenic positron beam capable of significantly improved energy resolution. The results provide evidence of resonances associated with \emph{infrared-inactive} vibrational modes, indicating that positron-molecule bound states may be populated by nondipole interactions. The anticipated ingredients for a theoretical description of such interactions are discussed.  
\end{abstract}


\maketitle

Positron-matter interactions are important in a variety of contexts, including astrophysics, materials science, and medical imaging~\cite{AstroPAHs09,Gidley2006,PrinciplesofPET}. One process of interest is the formation of positron-molecule bound states~\cite{RMP2010}, which can be populated by positron capture in vibrational Feshbach resonances (VFR)~\cite{Gribakin2000,Gribakin2001}. These VFR occur at incident positron energies 
\begin{equation}
{\varepsilon_\nu = \omega_\nu - \varepsilon_b},  
\label{Eq:ResEnergy}
\end{equation}
where $\omega_\nu$ is the energy of the molecular vibrational mode $\nu$ excited at capture, and $\varepsilon_b$ is the positron binding energy. The formation of these bound states typically results in significant enhancement of the positron annihilation rate near the resonant energy $\varepsilon_\nu$.

The theoretical description of these resonant annihilation processes requires knowledge of the mechanism which couples the motion of the light particle (positron) to the slow and heavy nuclear framework~\footnote{A similar process of low-energy electron attachment to molecules is usually mediated by a resonant state on the anion potential curve, which has no analog for positrons. On the other hand, VFR are observed in electron attachment to molecular clusters and in collisions with molecules, such as the hydrogen halides, see the review~\cite{Hotop2003}.}. The currently accepted theory relies on long-range dipole coupling between the positron and target molecule and therefore only describes VFR associated with infrared-active (IA) vibrational modes~\cite{Gribakin2006}. In general, vibrational excitations in low-energy positron (or electron) collisions require strong long-range interactions that depend sensitively on the internuclear separation~\cite{Lane80}. Studies of positron vibrational excitation of simple molecules (H$_2$, CO, CO$_2$, CH$_4$, CF$_4$~\cite{Sullivan2001,Sullivan2002,Marler2005}, and N$_2$~\cite{Gianturco1997}) show that, for IA modes, the cross sections are at least an order of magnitude greater than for those without dipole coupling. Thus, it is of significant interest to determine whether nondipole coupling mechanisms exist that would enable infrared-\emph{inactive} mode contributions to the positron-molecule annihilation spectrum.

Previously, measurements of the positron annihilation rate as a function of incident positron energy have shown evidence of VFR (and therefore positron-molecule bound states) for many molecules~\cite{RMP2010}. However, due to limitations in positron-beam energy resolution, fully resolved, individual VFR have only rarely been observed~\footnote{One of the few exceptions is CS$_2$, which has only three modes, producing one observable VFR at $\varepsilon \approx 0.12$~eV~\cite{Danielson2010}.}. To this end, positron-cooling and beam-formation processes were investigated~\cite{Natisin2014,Natisin2015,Natisin2016}, and a technique was developed to produce positron beams with significantly higher energy resolution than was available previously~\cite{Natisin2016b}. 

Presented here are measurements of positron annihilation made using this high-energy-resolution, cryogenic, trap-based positron beam. While positrons have been found to bind a wide range of polyatomic molecules, both polar and nonpolar \cite{RMP2010,Danielson2012a,Danielson2012b}, the two molecules studied here, 1,2-trans-dichloroethylene (C$_2$H$_2$Cl$_2$) and tetrachloroethylene (C$_2$Cl$_4$), were chosen specifically to investigate the possibility of positron capture in VFR mediated by nondipole interactions. Both molecules contain relatively well-isolated infrared-inactive vibrational modes, making them ideal candidates for this investigation.

The experimental apparatus and procedures for producing the cryogenic positron beam have been described in detail elsewhere~\cite{Natisin2016b}. Positrons emitted from a $^{22}$Na radioactive source are slowed to electronvolt energies using a layer of solid Ne maintained at 8~K~\cite{Greaves1996}. This steady-state beam is magnetically guided into a three-stage buffer-gas trap, which consists of a modified Penning-Malmberg (PM) trap in a ${\sim }0.1$~T magnetic field. The positrons are trapped and cooled through rotational excitation of a $300$~K N$_2$ buffer gas. After the positrons have been cooled for $0.1$~s they are ejected as a pulsed beam and retrapped in the recently developed cryogenic beam-tailoring trap (CBT). 

The CBT is a PM trap which is cryogenically cooled to $50$~K in a ${\sim}65$~mT magnetic field. Positrons trapped in the CBT are compressed radially using azimuthally rotating electric fields~\cite{Greaves2000,Greaves2008,Charlton2011} and axially by pulling them into a deeper potential well. The positrons are then cooled for $0.2$~s through vibrational and rotational excitation of the $50$~K CO buffer gas and subsequently ejected in pulses at a rate of ${\sim}1$~Hz with a total energy spread of $\Delta E_\text{tot} \sim 7$~meV FWHM. 

The CBT beam is passed through a gas cell containing the target molecular gas, where a CsI crystal detects single annihilation gamma rays during a 10~$\mu$s window as a function of the cell retarding potential. This process allows the annihilation rate to be measured as a function of the incident positron energy. By convention, the measured annihilation rate $\lambda$ is normalized by the Dirac annihilation rate $\lambda_\text{D}$, where $\lambda_\text{D}$ is the annihilation rate with a free electron gas with density $n$ equal to that of the target gas, yielding the dimensionless quantity $\Zeff$~\cite{RMP2010}:
\begin{equation}
\Zeff \equiv \frac{\lambda}{\lambda_\text{D}} = \frac{\lambda}{\pi r_0^2 c n}.
\label{Eq:Zeff}
\end{equation}
Here, $r_0$ is the classical electron radius, and $c$ is the speed of light.

For reasons not presently understood, the measured energy resolution of the beam in the gas cell, $\Delta E_\text{tot} \sim 20$~meV FWHM, is larger than the $\sim 7$~meV FWHM spread at the exit of the CBT. Nevertheless, the present measurements still represent more than a factor of two improvement over previous experiments.

The currently accepted theoretical description of energy-resolved positron annihilation in molecules was developed in Refs.~\cite{Gribakin2000,Gribakin2001,Gribakin2006,Gribakin2009}. The combined effect of the various annihilation mechanisms may be described as
\begin{equation}
\ZeffTot (\varepsilon ) = \ZeffDir (\varepsilon ) + \ZeffRes (\varepsilon ) + \eta \ZeffMRA (\varepsilon ).
\label{Eq:ZeffTot}
\end{equation}
The first term, $\ZeffDir$, describes direct ``in-flight'' annihilation and is typically small. The last term, $\ZeffMRA$, describes multimode resonant annihilation (MRA), which occurs due to direct positron capture into multimode vibrational states (i.e., overtones and combinations of the fundamentals)~\cite{Gribakin2009}. While the energy dependence predicted by $\ZeffMRA$ is typically in good agreement with measurements, its magnitude in small-sized polyatomics is often too large. Therefore, the $\ZeffMRA$ term is scaled by a constant factor $\eta$ to best fit the measured data~\cite{Jones2012,Natisin_thesis}.

For the work described here, the most important term in Eq.~\eqref{Eq:ZeffTot} is the second term, $\ZeffRes$. This term describes the contribution of individual resonances due to the excitation of fundamental molecular vibrations. In this case, $\ZeffRes$ may be written as a sum over the fundamental vibrational modes:
\begin{equation}
\ZeffRes \left( \varepsilon \right) = \pi F \sum_\nu g_\nu \sqrt{\frac{\varepsilon_b}{\varepsilon_\nu}} \frac{\Gamma^e_\nu }{\Gamma_\nu} f \left( \varepsilon_\nu - \varepsilon \right).
\label{Eq:ZeffRes}
\end{equation}
Here $g_\nu$ is the mode degeneracy, $\Gamma^e_\nu /\Gamma_\nu $ is the ratio of the elastic and total widths of resonance $\nu $, $f(\varepsilon_\nu - \varepsilon)$ is the positron energy distribution, and $F \approx 18$~eV (0.66~a.u.) describes the positron-electron overlap in the bound state~\cite{Gribakin2006,RMP2010,Gribakin2001}. For IA modes, positron capture is mediated by long-range dipole coupling, and the factor $\Gamma^e_\nu /\Gamma_\nu $ is easy to calculate
\cite{Gribakin2006}.

For all but the weakest IA modes, the positron annihilation rate in the VFR is significantly smaller than the detachment rate due to vibrational deexcitation. In this case $\Gamma^e_\nu /\Gamma_\nu \approx 1$, and the contributions of each VFR in Eq.~\eqref{Eq:ZeffRes} are simply set by $\sqrt{ \varepsilon_b / \varepsilon_\nu}$. In many molecules, however, the magnitudes of the measured VFR differ significantly from that predicted by Eq.~\eqref{Eq:ZeffRes}~\cite{RMP2010}. It is believed that this is due to a process known as intramolecular vibrational redistribution (IVR), in which the vibrational energy of the molecule is redistributed into near-resonant multimode states. In some circumstances, a mode-scaling factor $\beta_\nu$ is included in Eq.~\eqref{Eq:ZeffRes} to allow the magnitudes of the VFR to be fit to the measured data, thus quantifying the effects of IVR~\cite{Jones2013}. This factor is not used here in favor of displaying the unscaled predictions of the theory.

\begin{figure}
\includegraphics[width=\columnwidth,clip=]{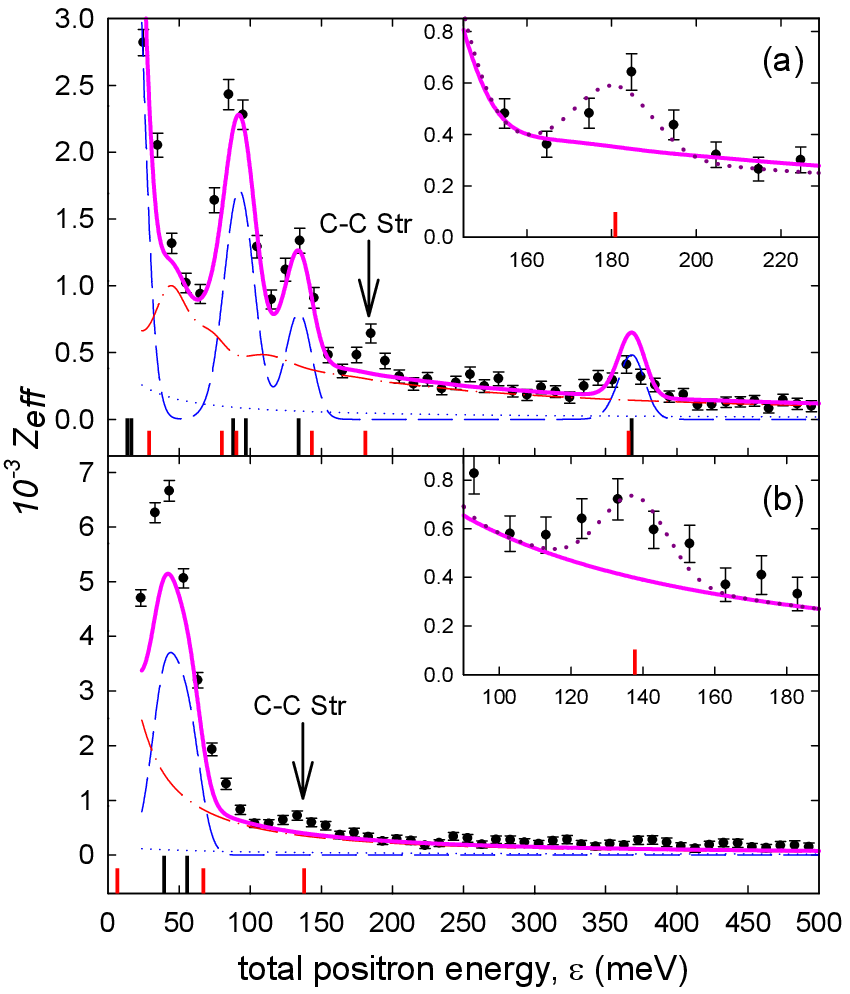}
\caption{Annihilation spectra for (a) 1,2-trans-dichloroethylene (C$_2$H$_2$Cl$_2$) and (b) tetrachloroethylene (C$_2$Cl$_4$): ($\bullet$) measured data; (\textcolor{magenta}{$\textbf{---}$}) total $\ZeffTot$ from Eq.~\eqref{Eq:ZeffTot}; (\textcolor{blue}{$\cdots$}) $\ZeffDir$; (\textcolor{blue}{$\textbf{-- --}$}) $\ZeffRes$, described by Eq.~\eqref{Eq:ZeffRes} and including only IA modes,
using $\Gamma^e_\nu /\Gamma_\nu =1$; and ($-\cdot-$) $\ZeffMRA$, scaled by the factor $\eta$. Vertical bars show resonant energies $\varepsilon_\nu$; tall black bars and short red bars denote IA and infrared-inactive modes, respectively. The fit parameters from Eq.~(\ref{Eq:ZeffTot}) are (a) $\varepsilon_b = 15$~meV, $\eta = 0.76$ and (b) $\varepsilon_b = 57$~meV, $\eta = 0.13$. Insets show unidentified resonances in detail, where (\textcolor{magenta}{$\cdots$}) show $\ZeffTot$ with $\ZeffRes$ obtained from Eq.~\eqref{Eq:ZeffRes} \emph{including} the contribution of infrared-inactive C-C stretch modes with $\Gamma^e_\nu /\Gamma_\nu $ chosen to give the best fit of the data, yielding $\Gamma^e_\nu /\Gamma_\nu =0.39$ and 0.22 in (a)  and (b), respectively.}
\label{Fig:AnnSpectra}
\end{figure}

Shown in Fig.~\ref{Fig:AnnSpectra} are the measured $\Zeff$, as defined in Eq.~\eqref{Eq:Zeff}. All error bars represent the standard error (1$\sigma$) associated with the statistical uncertainty in the measurements. Systematic uncertainties due to gas pressure and positron number are estimated to be ${\leq} 20~\%$ (not shown). The data shown are the average of several consecutive measurements which were done at two distinct gas pressures to ensure that $\Zeff$ is independant of gas pressure, as expected. Also shown are $\Zeff$ from Eq.~\eqref{Eq:ZeffTot}, including each of the terms, and using the vibrational mode energies from Ref.~\cite{Shimanouchi1972}.

Figure~\ref{Fig:AnnSpectra}(a) shows the averaged measured annihilation spectrum for 1,2-trans-dichloroethylene (C$_2$H$_2$Cl$_2$), where the measurements were made at gas pressures of approximately 6 and 13 $\mu$Torr. Several narrow spectral features may be seen that demonstrate the improved energy resolution of the CBT-based cryogenic positron beam. Due to the molecular symmetry, half of the vibrational modes are IA, while the other half are infrared inactive (distinguished by the colors and heights of the vertical bars along the bottom of the figure). Fitting Eq.~\eqref{Eq:ZeffTot} to the data for 1,2-trans-dichloroethylene yields a binding energy $\varepsilon_b = 15$~meV and an MRA scale factor $\eta = 0.76$. Here it is seen that $\ZeffTot$ from Eq.~\eqref{Eq:ZeffTot} is in relatively good agreement with the measured data for much of the spectrum, even though only the resonances of IA modes, for which we set $\Gamma^e_\nu /\Gamma_\nu =1$, are included in Eq.~\eqref{Eq:ZeffRes}. This suggests that the contribution from infrared-inactive modes is small. However, as discussed above, VFR magnitudes can deviate significantly from the model predictions for all but the simplest molecules (e.g., methyl halides), with magnitudes observed both above and below model predictions due to the effects of IVR~\cite{Jones2013}. This means that the magnitudes of the observed VFR are not expected to be reliable indicators of the effects of infrared-inactive mode contributions when the latter are near-degenerate with resonances of IA modes.

Of significant interest in Fig.~\ref{Fig:AnnSpectra}(a), however, is the resonance observed at $\varepsilon_\nu \approx 185$~meV (see inset). If this were due to a VFR, then from Eq.~\eqref{Eq:ResEnergy}, accounting for the 15~meV binding energy, this resonance would be given by a mode with energy $\omega_\nu \approx 200$~meV. This is quite close to the energy of the infrared-inactive C-C stretch mode with $\omega_\nu = 196$~meV~\cite{Shimanouchi1972}. This dipole-inactive mode has $A_g$ symmetry; thus, as discussed below, it could be excited either through a quadrupole interaction or through polarization or short-range interactions. As a comparison, the maximum deviation between the peak of the observed IA resonances and their expected energies (i.e., the $\epsilon_\nu$ given by Eq.~\ref{Eq:ResEnergy}) is 3~meV, which is comparable to the 4~meV deviation between the unidentified resonance and that expected from the C-C stretch mode. For reference, the contribution for a mode with energy $\omega_\nu = 196$~meV is shown in the inset to Fig.~\ref{Fig:AnnSpectra}(a), obtained by using Eq.~\eqref{Eq:ZeffRes} with $\Gamma^e_\nu /\Gamma_\nu = 0.39$, fitted to reproduce the magnitude of the measured $\Zeff$.

As a second example, the average measured annihilation spectrum for tetrachloroethylene (C$_2$Cl$_4$) is shown in Fig.~\ref{Fig:AnnSpectra}(b), measured at gas pressures of approximately 6 and 11 $\mu$Torr. Fitting the measured data to Eq.~\eqref{Eq:ZeffTot} yields a binding energy $\varepsilon_b = 57$~meV and an MRA scale factor $\eta = 0.13$. In this case, seven of the twelve modes are infrared inactive, and many of the resonances are shifted below zero energy due to the relatively high binding energy. Here, the magnitude of the low-energy resonance is ${\sim} 25\%$ larger than that predicted by Eq.~\eqref{Eq:ZeffTot} using only IA-mode contributions in Eq.~\eqref{Eq:ZeffRes}, though again, the possibility of IVR complicates the analysis.  

As in the first example, Fig.~\ref{Fig:AnnSpectra}(b) shows an isolated resonance in the measured data which is not accounted for by Eq.~\eqref{Eq:ZeffRes}. In this case, the unidentified resonance occurs at $\varepsilon_\nu \approx 135$~meV [see Fig.~\ref{Fig:AnnSpectra}(b) inset], which, from Eq.~\eqref{Eq:ResEnergy}, would be a VFR from a mode with energy $\omega_\nu \approx 192$~meV. As above, this energy is close to the infrared-inactive C-C stretch at $\omega_\nu = 195$~meV (also $A_g$ symmetry). In this case, the maximum deviation between the expected and observed IA mode energies is 4~meV, which is again comparable to the 3~meV deviation between the unidentified resonance and the expected infrared-inactive VFR. As shown on the inset in Fig.~\ref{Fig:AnnSpectra}(b), its contribution can be described by Eq.~\eqref{Eq:ZeffRes} for a mode with energy $\omega_\nu = 195$~meV by using $\Gamma^e_\nu /\Gamma_\nu = 0.22$, fitted to reproduce the magnitude of the measured $\Zeff$.

\begin{figure}
\includegraphics[width=\columnwidth,clip=]{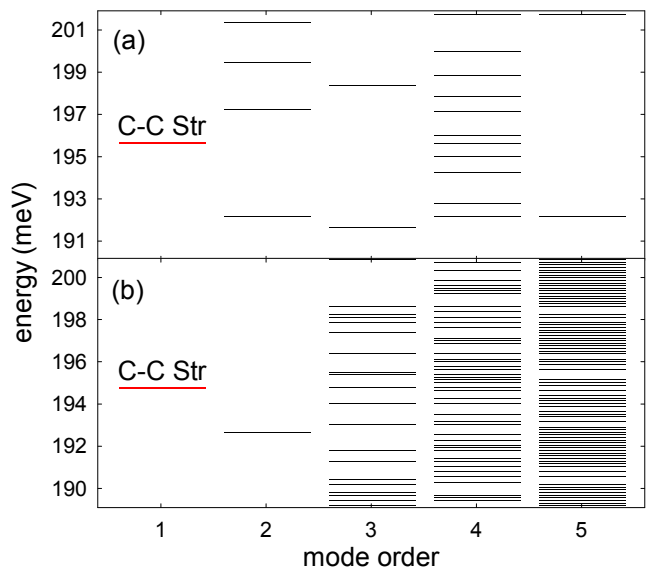}
\caption{Energies of multimode vibrational states within 5~meV of the C-C stretch up to mode order 5 for (a) 1,2-trans-dichloroethylene, and (b) tetrachloroethylene.}
\label{Fig:ModePlots}
\end{figure}

These two examples demonstrate what appears to be VFR populated through nondipole interactions; however, it is worth considering other possible explanations.  For example, a question could be raised as to whether the new features could be due to IR active modes of impurities present in the sample gases. This is unlikely, since positron binding energy is highly molecule specific \cite{Danielson2012a} and is unlikely to coincide within a few millivolts of the molecules studied. Another alternative is direct excitation of a multimode state. To explore this possibility, Fig.~\ref{Fig:ModePlots} shows the vibrational multimode spectrum, in the harmonic approximation, up to mode order 5 within 5~meV of the C-C stretch mode for both 1,2-trans-dichloroethylene [Fig.~\ref{Fig:ModePlots}(a)] and tetrachloroethylene [Fig.~\ref{Fig:ModePlots}(b)], where mode order represents the number of constituent fundamentals in a given multimode state.

As seen in Fig.~\ref{Fig:ModePlots}, there are several multimodes near the C-C stretch mode in both molecules. The process of direct excitation of multimode states is described quantitatively by the $\ZeffMRA$ term in Eq.~\eqref{Eq:ZeffTot} and has been examined in detail elsewhere~\cite{Jones2012}. The MRA model is typically found to over-predict the spectral weight due to multimode annihilation in small-to-medium-sized polyatomics, but it provides reasonable agreement with the measured data once scaled by a single numerical factor. This is seen in both examples shown in Fig.~\ref{Fig:AnnSpectra}, where the scaled MRA contribution describes well the smooth background in the regions between the VFRs. However, the fact that the model does not predict any clear resonances near those observed in the measured data suggests that direct excitation of multimodes is an unlikely explanation of the observed features.  

It is possible, however, that the observed resonances could be due to a combination of direct multimode excitation and IVR, which is not accounted for in the MRA model. For this mechanism, the positron would need to excite a particular IA multimode vibration and couple through IVR into nearby multimode states with a slower positron detachment rate than the entrance state. This process would result in a longer positron dwell time on the molecule, and therefore a larger annihilation rate, than predicted by the MRA model (see, e.g., Ref.~\cite{Danielson2013}). However, this explanation for the observed isolated resonances appears unlikely. There is a dense ``background'' of multimode states in virtually all molecules studied to date in which this process could potentially occur. It would therefore be highly coincidental for the unidentified resonances to occur at precisely the two locations where an infrared-inactive mode is expected to contribute.

Thus, the data presented here display two examples of well-isolated, fully resolved VFR in the measured annihilation spectra that cannot be described using existing theory. Further, both resonances occur at energies that correspond to those expected for infrared-inactive vibrational excitations. To include the contribution of such resonances to $\ZeffRes$ would require a significant extension of current theory. 

To describe the contribution of an isolated VFR
using Eq.~\eqref{Eq:ZeffRes}, one needs to evaluate its capture (or elastic) width $\Gamma^e_\nu$ and total width $\Gamma _\nu =\Gamma ^e_\nu +\Gamma ^a$, where $\Gamma^a \approx 0.03 \sqrt{\varepsilon_b[\text{meV}]}~\mu$eV (i.e., $1.1\times 10^{-9}\sqrt{\varepsilon _b[\text{meV}]}$~a.u.) is the annihilation width of the positron bound state~\cite{Gribakin2001,Gribakin2006,RMP2010}. In general, $\Gamma_\nu$ can also contain contributions due to vibrationally inelastic positron escape~\cite{RMP2010,Jones2013,Danielson2013}. This effect occurs due to mode mixing and can be assumed to be relatively weak in small polyatomics. Also, as mentioned above, the smallness of $\Gamma^a$ means that even a small coupling (e.g., that of a weak IA mode) can support a ``full-sized'' VFR with $\Gamma^e_\nu /\Gamma_\nu\approx 1$, which simplifies the application of Eq.~\eqref{Eq:ZeffRes}.

For IA modes the elastic VFR widths are determined by the corresponding vibrational transition dipole amplitudes~\cite{Gribakin2006} and have typical
values $\Gamma^e_\nu \sim 1$--$10~\mu$eV ($10^{-7}$--$10^{-6}$~a.u.) (e.g., for methyl halides). For infrared-inactive modes the possible coupling mechanisms may involve positron interaction with the molecular quadrupole moment or polarization potential, or some short-range interactions.
In the former case we can use the approach of Ref.~\cite{Gribakin2006} to estimate the elastic width,
\begin{equation}
\Gamma^e_\nu =\frac{64}{(15)^3}\omega_\nu^2|Q_\nu|^2g(\xi ),
\label{eq:Q}
\end{equation}
where $Q_\nu $ is the quadrupole transition amplitude for the excitation of mode $\nu$, and $g(\xi )$ is a dimensionless function of $\xi =1-\varepsilon _b/\omega _\nu$, such that $h(0)=h(1)=0$, and $g_{\rm max}\approx 0.883$ at $\xi \approx 0.935$~\footnote{Explicitly, ${g(\xi )=\xi ^{5/2}(1-\xi )^{-1/2} \left[ _2F_1\left(\frac{1}{2},1;\frac{7}{2};-\xi /(1-\xi )\right) \right]^2}$.}. Compared with the corresponding expression for the dipole-driven (IA) modes [Ref.~\cite{Gribakin2006}, Eq.~(7)], Eq.~\eqref{eq:Q} contains an extra power of $\omega_\nu$, which can suppress positron capture by infrared-inactive modes. From the fits shown on the insets of Fig.~\ref{Fig:AnnSpectra} (a) and (b), the magnitudes of the non-dipole resonances are consistent with C-C stretch resonances described by Eq.~\eqref{Eq:ZeffRes} with $\Gamma^e_\nu /\Gamma_\nu =0.39$ and 0.22 for C$_2$H$_2$Cl$_2$ and C$_2$Cl$_4$, respectively. From these values we can determine $\Gamma^e_\nu$ and use Eq.~\eqref{eq:Q} to find the corresponding values of the transition amplitudes, yielding $Q_\nu =0.056$ and 0.072~a.u. These values are close to known values of quadrupole transition amplitudes \footnote{There are few calculations of vibrational quadrupole transition amplitudes. Values of $Q_\nu$ are available for H$_2$ (0.088~a.u. \cite{Poll1978}) and N$_2$ (0.059~a.u. \cite{Li2007}), and can be estimated for the symmetric stretches in CO$_2$ (0.033~a.u.) and C$_2$H$_2$ (0.043 and 0.017~a.u. for C-H and C-C, respectively) using data from Refs. \cite{Lindh1991,Gianturco1996}.}, which indicates that the quadrupole coupling can
contribute to positron capture by infrared-inactive modes.

Another long-range coupling mechanism is through the positron-molecule polarization potential $-\alpha ({\bf R})/2r^4$. Here $\alpha ({\bf R})$ is the molecular dipole polarizability, which depends on the nuclear coordinates ${\bf R}$. There is also an anisotropic quadrupole polarizability term, which is important for rotational excitations of molecules~\cite{Lane80}. Due to the strong singularity at small $r$, it is more difficult to estimate the corresponding elastic width. The transition amplitude will depend sensitively on how the polarization potential is cut off at small distances, and on the potential and the positron wave functions at short range. Also, the spherical part of the polarization potential can couple the positron partial wave of the highest symmetry (i.e., $s$-wave type) to the positron bound state of the same symmetry. Accurate calculation of this amplitude would require knowledge of both the continuum- and bound-state wave functions that are strongly affected by the positron-molecule correlation potential (which provides binding). The associated difficulties are similar to those that have hampered calculations of positron binding to nonpolar molecules using standard quantum-chemistry techniques~\cite{RMP2010}.

Presented here are measurements of positron annihilation using a recently developed, high-energy-resolution, cryogenic positron beam. Data from two molecules are described in which well-isolated resonances are observed that cannot be explained on the basis of existing theory. In both cases, these resonances occur at energies consistent with the excitation of infrared-inactive vibrational modes, providing strong evidence that positron-molecule bound states may be populated via nondipole interactions. Although only two examples were shown here, this effect is likely to be important in a wide variety of molecules. These results emphasize the need for a quantitative theoretical description of VFR mediated by short-range and nondipole interactions. As a first step towards this end, the anticipated elements of such a theory were discussed.  

This work is supported by NSF, grant PHY 14-01794.

\bibliography{NonDipoleVFRs}

\end{document}